\documentstyle[11pt,newpasp,twoside]{article}
\def \lya{Ly-$\alpha$}

\def \hi{H$\,\sc i$}

\def \hd{HD$\,$209458b}
\def\ga{\mathrel{\mathchoice {\vcenter{\offinterlineskip\halign{\hfil
$\displaystyle##$\hfil\cr>\cr\sim\cr}}}
{\vcenter{\offinterlineskip\halign{\hfil$\textstyle##$\hfil\cr>\cr\sim\cr}}}
{\vcenter{\offinterlineskip\halign{\hfil$\scriptstyle##$\hfil\cr>\cr\sim\cr}}}
{\vcenter{\offinterlineskip\halign{\hfil$\scriptscriptstyle##$\hfil
\cr>\cr\sim\cr}}}}}

\def\edcomment#1{\iffalse\marginpar{\raggedright\sl#1\/}\else\relax\fi}
\reversemarginpar

\begin{document}
\title{Evaporation rate of hot Jupiters and formation of Chthonian planets}
 \author{G. H\'ebrard, A. Lecavelier des \'Etangs, A. Vidal-Madjar, 
J.-M. D\'esert \& R. Ferlet}
\affil{Institut d'Astrophysique de Paris, CNRS, 98$^{\rm bis}$ boulevard 
Arago, F-75014 Paris, France}

\begin{abstract}
Among the hundred of known extrasolar planets, about 15\% 
are closer than 0.1~AU from their parent stars. But there are
extremely few detections of planets orbiting in less than 3
days. At this limit the planet \hd\ has been found to have an
extended upper atmosphere of escaping hydrogen. This suggests that the
so-called hot Jupiters which are close to their parent stars could
evaporate.

Here we estimate the evaporation rate of hydrogen from
extrasolar planets in the star vicinity. With high exospheric
temperatures, and owing to the tidal forces, planets evaporate through
a geometrical blow-off. This may explain the absence of Jupiter mass
planets below a critical distance from the stars. Below this critical
distance, we infer the existence of a new class of planets made of 
the residual central core of former hot Jupiters, which we propose to
call the ``Chthonian'' planets.
\end{abstract}

Following the recent discovery of a significant escape of atomic
hydrogen from the planet orbiting HD$\,$209458 (Vidal-Madjar et
al.~2003), we propose to evaluate the escape flux from the upper
atmosphere of hot Jupiters under the influence of the strong tidal
forces from their parent stars. This escape flux had previously been
estimated (Guillot et al.~1996), but the conclusion that the mass loss
is not significant was based on two hypotheses which need
to~be~revisited.

\vspace{0.2cm}
First, the black body radiative equilibrium used to calculate the
temperature of the upper atmosphere (Schneider et al.~1998) is
inappropriate because it does not apply to the low density upper
atmosphere. Observations in the Solar System show that the temperature
of a planetary upper atmosphere (thermosphere, exosphere) is much
higher than the effective temperature of the bottom atmosphere.  For
example, whereas the temperatures in the Earth and Jupiter are around
200~K and 120~K respectively at the level of the tropopause, they
reach 1000~K in the thermosphere of these two planets.  Although these
observed high temperatures in the planets of our own Solar System
remain unexplained, there are some clues that a combination of the
extreme and far ultraviolet fluxes with the Solar wind is responsible
for the~heating.

The second important hypothesis is tidal force, which modifies the
gravity of hot Jupiters. The common hypothesis is to neglect that
effect as for isolated planets far from their star. However, tidal
forces have a significant influence on the density distribution in the
upper atmosphere of hot Jupiters.

\vspace{0.2cm}
At a given temperature the escape flux can be calculated using the
Jeans' escape estimate. {\it Jeans' escape} refers to the escape of
atoms and molecules whose velocities are in the tail of the Boltzmann
distribution and which have enough energy to escape the planet
gravity. The flux is calculated at the top of the thermosphere, that
is the level above which the atoms and molecules have no
collisions. The exobase is usually defined as the location above which
the mean free path is larger than the scale height of the
atmosphere. Here we extend this idea by defining the exobase as the
location above which the mean free path is larger than the distance to
the Roche lobe of the planet.  Thus, the exobase is the location above
which the atoms and molecules can definitively escape the~planet.

We computed the escape flux of atomic and molecular hydrogen as a
function of the upper atmosphere temperature.  From this, we can
derive the corresponding life time needed to evaporate the total mass
of the planet.  For $T\ga7000$\,K, the \hi\ escape flux is larger than
the minimum flux of $10^{10}$\,g\,s$^{-1}$ determined from \lya\
observations (Vidal-Madjar et al.~2003).

\vspace{0.2cm}
Although the mechanism responsible for the heating of the upper
atmosphere of the planets in the Solar System is not fully identified,
we can hint a simple estimate of a plausible temperature from the
comparison of the heating and cooling mechanisms. A lower limit of the
heating can be estimated from the energy flux of the stellar extreme
ultraviolet and the Ly-$\alpha$ photons.  The cooling is due to a
combination of the heat conduction toward the cooler bottom
atmosphere, the collisional ionization and excitation of the \hi\ 
electronic levels, and the evaporation at the top of the upper
atmosphere. Thus, a lower limit to the temperature can be estimated by
the energy balance.  With the parameters of \hd, we evaluate the 
temperature in the upper atmosphere to be at least
$\sim10000$\,K. With this temperature the present evaporation rate of
hydrogen from \hd\ must be $\sim 1$ to $5 \times 10^{11}$~g$\,$s$^{-1}$. 
This is in agreement with the lower limit found from the HST
Ly-$\alpha$ observations. Assuming an age of $5 \times 10^9$ years,
\hd\ may have lost about 1\% to 7\% of its mass.

\vspace{0.2cm}
Finally, we also estimated the life time of a given planet as a
function of its mass and orbital distance.  We found that planets with
orbital periods shorter than 2 to 3 days have very short life
time. This may explain that only few planets have been detected with
periods shorter than 3 days. Moreover, low-mass planets have also
short life time. This means that the nature of these low-mass hot
Jupiters should evolve with time. These planets should loose a large
fraction of their hydrogen.  This process may lead to planets with an
hydrogen-poor atmosphere, or even with no more atmosphere at all. The
emergence of the inside core of former and evaporated hot Jupiter may
give rise to a new class of planets which we proposed to call
``Chthonian'' planets in reference to the Greek god Khth\^on
(``Chthonian'' is used to name the Greek deities who come from hot
infernal underground).

\vspace{0.4cm}
\noindent
{\bf References:}

\noindent
Guillot, T., Burrows, A., Hubbard, W.~B., et al. 
1996, ApJL, 459, L35 

\noindent
Schneider, J., Rauer, H., et al.
1998, 
ASP Conf. Ser., 134, 241

\noindent
Vidal-Madjar, A., 
et al.
2003, Nature, 422, 143 (and this proceedings)

\vspace{0.3cm}
\noindent
{\small We thank J.\,C. McConnell, C.\,D. Parkinson, G.\,E. Ballester \& 
M. Mayor for~their~help.}

\end{document}